\documentclass[10pt,twocolumn]{article} 

\usepackage{ol2}
\usepackage[draft]{hyperref}
\usepackage{graphicx}
\usepackage{epsfig}
\usepackage{psfrag}
\usepackage{amssymb}
\usepackage{amsmath}

\newcommand{\rg}{{\bf r}}

\newcommand{\GO}{\Gamma_0}
\newcommand{\GR}{\Gamma_\mathrm{R}}
\newcommand{\GNR}{\Gamma_{\mathrm{NR}}}
\newcommand{\pg}{{\bf p}}
\newcommand{\Eg}{{\bf E}}
\newcommand{\Gg}{{\bf G}}

\begin{document}

\twocolumn[ 

\title{Distance dependence of the local density of states in the near field \\ of a disordered plasmonic film}


\author{E. Castani\'e, V. Krachmalnicoff, A. Caz\'e, R. Pierrat, Y. De Wilde and R. Carminati$^{*}$}

\address{Institut Langevin, ESPCI ParisTech, CNRS, 10 rue Vauquelin,
75231 Paris Cedex 05, France

$^*$Corresponding author: remi.carminati@espci.fr
}

\begin{abstract}
We measure the statistical distribution of the photonic local density of states in the near field of a semi-continuous gold film. 
By varying the distance between the measurement plane and the film, we show that near-field confined modes play
a major role in the width of the distribution. Numerical simulations in good agreement with experiments allow us to point out the influence of non-radiative decay channels at short distance.
\end{abstract}


 ] 

\noindent

The local density of optical states (LDOS) is a quantity of fundamental interest for the understanding and control of light-matter interaction
in structured environments~\cite{CPS,QED,NovotnyBook}. Plasmonic structures are expected to produce substantial changes in the optical modes
both in space and frequency, offering degrees of freedom for LDOS engineering on the nanometer scale at visible and near infrared 
wavelengths. The LDOS can be probed using fluorescent emitters~\cite{vanHulst2009,Krachmalnicoff2010,Frimmer2011} or thermal 
radiation~\cite{Joulain2003,DeWilde2006}. Qualitative LDOS mapping can also be performed in scattering experiments using angularly-integrated incoherent illumination or detection~\cite{Chicanne2002,Bouhelier2008}.

The LDOS $\rho(\rg_0,\omega)$ at position $\rg_0$ and frequency $\omega$ is
\begin{equation}
\rho(\rg_0,\omega) = \frac{2\omega}{\pi c^2} \mathrm{Im} \left [ \operatorname{Tr} \, \Gg(\rg_0,\rg_0,\omega) \right ]
\label{eq:defLDOS}
\end{equation}
where $\Gg$ is the dyadic Green function, such that  the electric field radiated at point $\rg$ by a point electric dipole $\pg$ located 
at point $\rg_0$ reads $\Eg(\rg)=\mu_0 \omega^2 \Gg(\rg,\rg_0,\omega) \pg$. This LDOS is relevant for the spontaneous emission
by an electric dipole emitter with position $\rg_0$ and emission frequency $\omega$, and without defined orientation~\cite{NovotnyBook}.
The decay rate averaged over the orientations of the transition dipole $\pg$ reads:
\begin{equation}
\Gamma = \frac{\pi \omega}{3 \epsilon_0 \hbar} |\pg|^2 \rho(\rg_0,\omega) \ .
\label{eq:defGamma}
\end{equation}
Measuring the fluorescence lifetime $\tau=1/\Gamma$ is a direct way to probe the LDOS. In plasmonics,
the seminal work of Drexhage demonstrated substantial changes in the lifetime when approaching a flat silver surface~\cite{CPS,Drexhage1968}.
More generally, the distance dependence of fluorescence signals in the near field of flat surfaces~\cite{Vahid05}, isolated 
nanoparticles~\cite{Carminati06,Novotny06} or nanoparticles arrays~\cite{Viste2010}
reveals the interplay between radiative and non-radiative processes. Understanding this interplay is of high interest for the design
of efficient light sources (radiative processes), or for the control of fluorescence quenching and nanoscale energy transfer (non-radiative processes).

In this Letter, we study the distance dependence of the LDOS in the near field of a disordered semi-continuous gold film. 
Such films are known to support unconventional plasmon modes, that, in particular, can localize optical energy in subwavelength areas
(hot spots)~\cite{Krachmalnicoff2010,Shalaev_review,Stockman2001,Gresillon1999,Laverdant2008}.
We measure statistical distributions of the LDOS at different distances in the near field, and compare the experimental data to numerical 
simulations. This allows us to assess the role of radiative and non-radiative channels.

\begin{figure}[htb]
\centerline{\includegraphics[width=8.3cm]{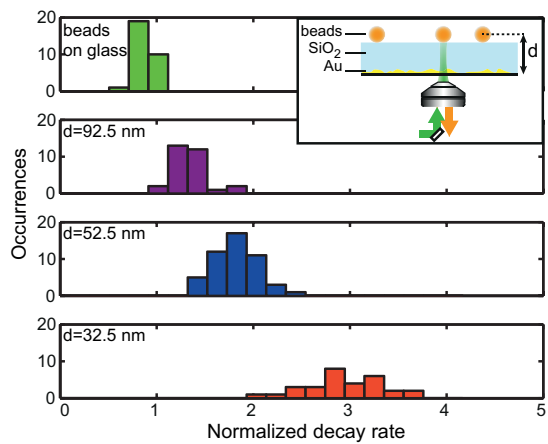}}
\caption{Distributions of the decay rate (normalized by its averaged value on a glass cover-slide) for different distances $d$ between
the center of the fluorescent nanosources and the film.  Inset: sketch of the setup.}
\label{exp}
\end{figure}

The experimental setup and the sample geometry are sketched in the inset in Fig.~\ref{exp}. 
The sample preparation consists in the following steps. First, gold is evaporated on a microscope cover-slide, in order to obtain a 
semi-continuous film around the percolation threshold. Second, a $\mathrm{SiO}_2$ spacer is evaporated on the gold film. 
Third, fluorescent nanosources (polystyrene beads, Invitrogen Fluospheres Red 580/605, diameter $25$ nm) are spin-coated on the top of the silica spacer. 
Several samples with different spacer thicknesses ranging from $20$ to $80$ nm have been produced.
 The experimental setup used to measure the fluorescence lifetime of the nanosources has been described previously~\cite{Krachmalnicoff2010}. 
We briefly recall the main features. The nanosources are individually illuminated through an oil immersion objective mounted on an inverted microscope. 
The excitation is performed at a repetition rate of 10~MHz by a picosecond super-continuum laser (Fianium SC450) filtered at the excitation wavelength 560~nm,
with an intensity $\simeq 10^3\;\mathrm{~W/cm}^2$. Fluorescence photons are collected through the same objective. They are separated from excitation photons 
by a dichroic mirror followed by a set of filters centered around $\lambda=607$~nm, with a width of $70$~nm, and are detected by an avalanche photodiode 
(Perkin Elmer SPCM-AQR-15). Lifetime measurements are performed by time-correlated single photon counting (PicoQuant TimeHarp200).

Figure~\ref{exp} displays the distributions of the decay rate at several distances $d$ between the center of the nanosources and the gold film, obtained from lifetime
measurements on 30 nanosources. One clearly
sees that both the mean value and the width of the distribution are affected by the proximity of the disordered gold film.
The behavior of the averaged value can be qualitatively understood replacing the semi-continuous film by an effective homogeneous film. 
The averaged LDOS close to this absorbing homogeneous film is expected to be larger than that on the glass cover-slide (our reference) 
due to non-radiative channels, and to increase at short distance~\cite{CPS}.
The broadening of the decay rate distribution when the distance to the film decreases is more interesting. The disordered nature of the film induces LDOS fluctuations, that are known to become substantial in the presence of near-field interactions able to generate large field variations on scales smaller than the wavelength~\cite{Krachmalnicoff2010,Sapienza2011}.
An important feature in our data is the substantial change in the shape and width of the statistical distribution with the distance to the film, in the range $d \simeq 30-90$~nm. This is due to spatial filtering of optical modes laterally confined on scales below the wavelength. Indeed, the field
distribution in a plane at a distance $d$ is exponentially filtered in Fourier space by a factor $\sim\exp(-K d)$ compared to the distribution at $d=0$~nm,
with $K$ the spatial frequency in the transverse direction (parallel to the film plane). This is a feature of near-field optics~\cite{NovotnyBook}. 
From this simple observation, an order of magnitude of the lateral confinement $\xi$ of the field can be extracted. Since field
variations giving rise to substantial fluctuations of the decay rate strongly attenuate between $d=32.5$~nm and $d= 52.5$~nm, 
the attenuation length can be estimated to be $1/K \simeq 10$~nm. One can deduce $\xi \sim 2\pi/K \simeq 60$~nm as a typical size
of hot spots, in agreement with orders of magnitude found by near-field optical microscopy~\cite{Gresillon1999,Laverdant2008}.

\begin{figure}[htb]
\centerline{\includegraphics[width=8cm]{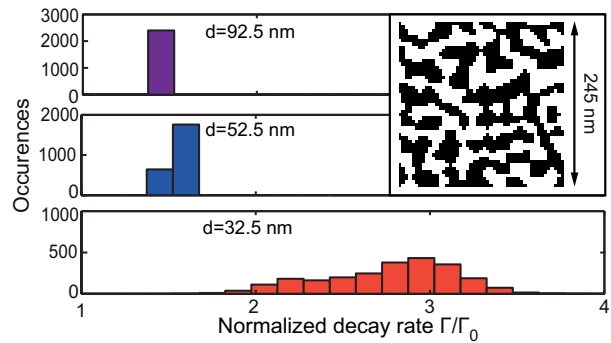}}
\caption{Calculated distributions of the normalized decay rate $\Gamma/\Gamma_0$ ($\Gamma_0$ is the value in vacuum) for different distances $d$, as in Fig.~\ref{exp}. $\lambda=605$~nm.
Inset: Film used in the simulation (black color corresponds to gold).}
\label{theo1}
\end{figure}

In order to have a better insight into the physical phenomena driving the decay rate distributions, we have performed numerical simulations. 
We have generated semi-continuous gold films using a Kinetic Monte-Carlo algorithm~\cite{Mottet98} that mimics the evaporation
and deposition  process. 
The principle is to let $5$~nm gold particles, interacting via a rescaled atomic potential taken from~\cite{Cleri93}, diffuse on a 2D square matrix until a 
stable configuration is reached. A typical realization of a film with feature comparable to that used experimentally is shown in the inset in Fig.~\ref{theo1}. 
Numerically generated films exhibit fractal clusters close to the percolation threshold, as observed in TEM images obtained on real films~\cite{Krachmalnicoff2010}.
To calculate the electric field at any position $\rg$ above the film, under illumination by a point electric dipole $\pg$ located at position $\rg_0$, we use a volume integral 
method based on the solution of the Lippmann-Schwinger equation:
\begin{equation}
\Eg(\rg) = \Eg_0(\rg) + \frac{\omega^2}{c^2} [\epsilon(\omega) - 1] \int_V \Gg_0(\rg,\rg^\prime,\omega) \Eg(\rg^\prime) \mathrm{d}^3 r^\prime
\end{equation}
where $\Gg_0$ is the vacuum dyadic Green function, $\epsilon(\omega)$ the dielectric function of gold, $V$ the volume
occupied by the gold film and $\Eg_0(\rg)=\mu_0 \omega^2  \Gg_0(\rg,\rg_0,\omega) \pg$ is the incident field. 
The numerical computation is done by a moment method, discretizing $V$ into $2.5$~nm cells. The Green function $\Gg_0$ is integrated over the cell to improve convergence~\cite{Chaumet04}. Once the electric field is known, the full Green function $\Gg$ is deduced
from $\Eg(\rg)=\mu_0 \omega^2 \Gg(\rg,\rg_0,\omega) \pg$, and the decay rate is computed from Eqs.~(\ref{eq:defLDOS}) and (\ref{eq:defGamma}).

\begin{figure}[htb]
\centerline{\includegraphics[width=8cm]{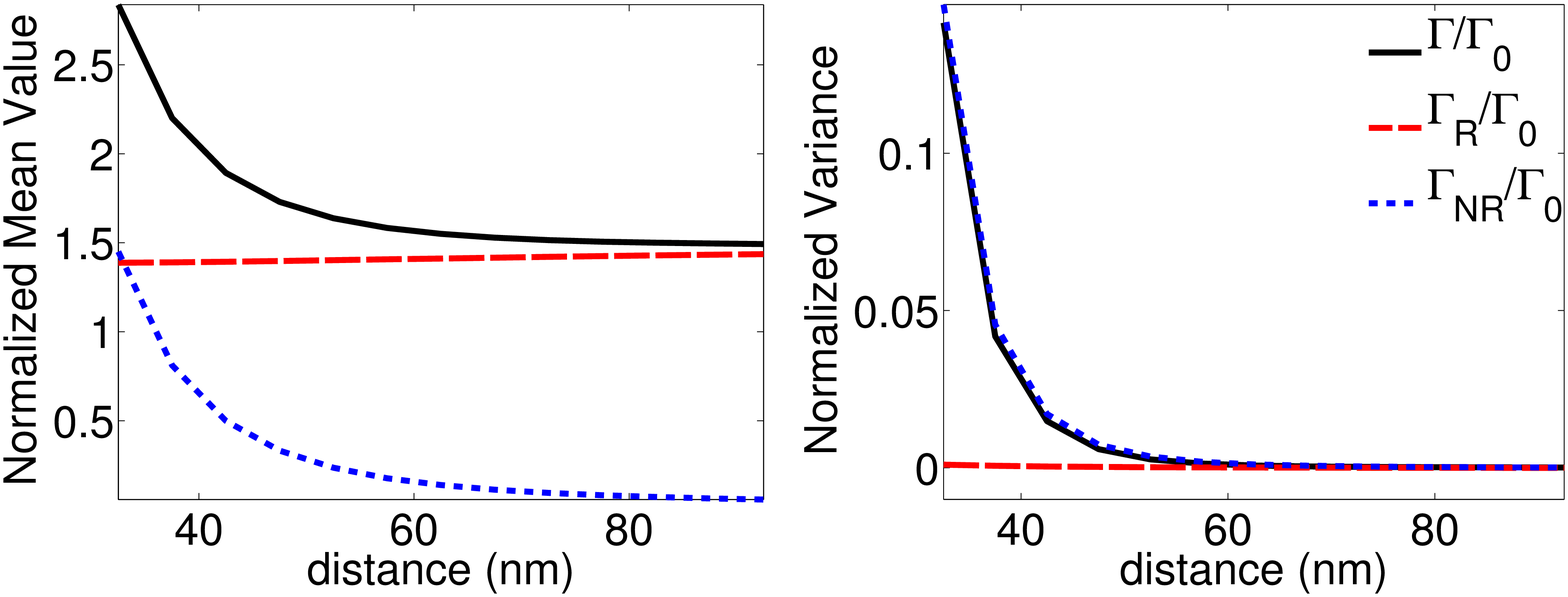}}
\caption{Mean value (left) and variance (right) of the distribution of $\Gamma/\Gamma_0$ (black solid line), $\GNR/\Gamma_0$ (blue dotted line) and $\GR/\Gamma_0$ (red dashed line) versus the distance $d$.}
\label{theo4}
\end{figure}

Calculated distributions of decay rates are shown in Fig.~\ref{theo1}, for the same distances $d$ as in Fig.~\ref{exp}.
The nanosources are modeled by orientationally averaged electric dipoles distributed inside spheres with diameter 25 nm. The metal film
and the nanosources are immersed in a homogeneous glass background that models the glass substrate and the silica spacer.
Experimental and numerical data show a very good qualitative agreement for the trends of both the mean and the width of the distributions. 
An interesting aspect of numerical simulations is the possibility of computing separately the radiative and non-radiative decay rates
$\GR$ and $\GNR$. The radiative decay rate is proportional to the power radiated in the far
field, while the non-radiative decay rate is proportional to the power absorbed in the sample~\cite{Carminati06,Vandembem2010}.
The mean value and variance of the distributions of $\Gamma/\Gamma_0$, $\GR/\Gamma_0$ and $\GNR/\Gamma_0$ are represented
 in Fig.~\ref{theo4} versus the distance $d$.
The increase of $\langle\Gamma\rangle$ at short distance is mainly due to the increase of $\langle\GNR\rangle$. 
This can be understood recalling the effective homogeneous film approach. Indeed, one does not expect significant variations
of $\langle\GR\rangle$ on such a distance range, since variations in the radiative rate are due to interference effects that build up on a length scale of the order of $\lambda/2 \gg d$. Interestingly, the broadening at short distance is also driven by non-radiative decay channels.
The confined near-field variations responsible for large near-field fluctuations of the LDOS are chiefly generated by non-radiative modes.

\begin{figure}[htb]
\centerline{\includegraphics[width=8cm]{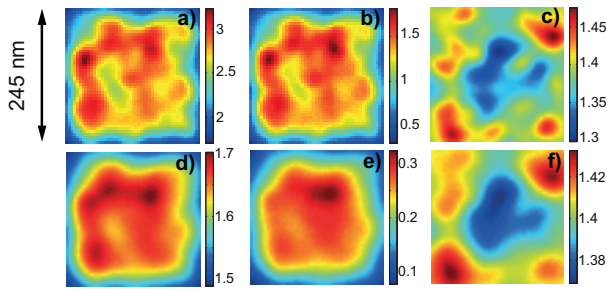}}
\caption{Maps of the  normalized decay rates for a single orientationally averaged dipole at a distance $d=32.5$~nm (a, b, c) and $d=52.5$~nm (d, e, f) above the film shown in Fig.~\ref{theo1}. (a, d): $\Gamma/\GO$, (b, e): $\GNR/\GO$, (c, f):  $\GR/\GO$.}
\label{theo2}
\end{figure}

The numerical calculation also allows us to compute spatial maps of both $\GR$ and $\GNR$ on a specific sample, as shown in Fig.~\ref{theo2}.
The similarity between the spatial distribution of $\Gamma$ and $\GNR$ is striking, confirming the influence of non-radiative
decay channels on the LDOS variations. The spatial distribution of $\GR$ and $\GNR$ 
do not coincide, showing that some specific points on the sample might give rise preferentially to a coupling to radiative or non-radiative channels. Finally, we point out that the spatial extent $\xi$ of bright LDOS spots is in agreement with the qualitative estimate 
based on the experimental data.

In summary, we have studied the distance dependence of the statistical distributions of the decay rate (LDOS) in the near field of a 
disordered semi-continuous gold film. Qualitative analyses and numerical simulations have allowed us to put forward the influence of
confined near-field modes, and to assess the role of non-radiative channels at short distance. This work illustrates the potential
of LDOS statistics in the study of optical modes in nanostructured disordered materials.


\pagebreak

\section*{Informational Fourth Page}
In this section, please provide full versions of citations to
assist reviewers and editors (OL publishes a short form of
citations) or any other information that would aid the peer-review
process.

\end{document}